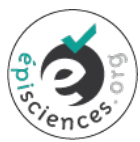
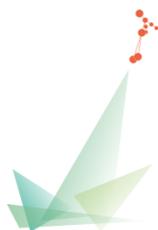
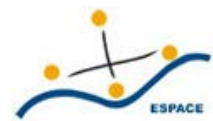

# Interstellar ices: a possible scenario for symmetry breaking of extraterrestrial chiral organic molecules of prebiotic interest

**Louis L.S. d'HENDECOURT\*[1,2], Paola MODICA[3], Cornelia MEINERT[4], Laurent NAHON[5], Uwe J. MEIERHENRICH[4]**

[1] IAS-CNRS-Université Paris-Sud, Campus d'Orsay, France
[2] PIIM-CNRS-Aix Marseille Université, Marseille, France
[3] LPC2E-CNRS-Université d'Orléans, Orléans, France
[4] ICN-CNRS-Université de Côte d'Azur, Nice, France
[5] Synchrotron SOLEIL, Gif-sur-Yvette, France

\*Corresponding author: ldh@ias.u-psud.fr

**Abstract**

In the laboratory, the photo- and thermochemical evolution of ices, made of simple molecules of astrophysical relevance, always leads to the formation of semi-refractory water-soluble organic residues. Targeted searches for specific molecules do reveal the notable presence of two families of important molecular "bricks of life": amino acids, key molecules in metabolism, and sugars, including ribose, the backbone of RNA molecules which support the genetic information in all living entities. Most of these molecules are indeed found in primitive carbonaceous meteorites and their implication in prebiotic chemistry at the surface of the early Earth must be seriously considered. These molecules are, almost all, chiral. In meteorites, some amino acids do show significant enantiomeric excesses, practically exclusively of the L-form. In our experiments, we investigate the role of circularly polarized light obtained from the DESIRS beamline of the synchrotron SOLEIL, a light commonly observed in regions of star formation, in order to generate an initial symmetry breaking in chiral amino acids produced and then indeed detected in our samples. We present first a brief global description of the chemical evolution of the Galaxy. Then, using our laboratory simulations, we suggest the importance of cosmic ices in the build-up of complex organic matter, including enantioenrichment at the surface of telluric planets like the Earth, thus establishing a link between astrochemistry and astrobiology.

# I ASTROPHYSICAL BACKGROUND

Interstellar ices are generally observed and well documented in molecular clouds surrounding protostars out of which stars, protoplanetary disks and planets but also many debris such as asteroids and comets do form (Öberg et al., 2014). It must be pointed out that, in molecular clouds, simple icy molecules from the most cosmically abundant elements (H, O, C, N, S and P), often dubbed as CHNOPS in the astrobiological community, constitute by far, the most abundant molecular species (Le Sergeant d'Hendecourt, 2011). Mid infrared vibrational spectroscopy is the only *direct* access to the chemical composition of interstellar grains and therefore to the composition of the ices (Dartois, 2005, Boogert et al, 2008). Radio astronomy provides for the observation of gas phase species. Slightly more than 200 molecules (ions, radicals and stable species) are indeed observed in the gas phase (see a good example in **http://www.pcmi.univ-montp2.fr/**), mostly toward dense clouds, but their overall abundances remain rather small as compared to molecular ices. The accretion of gas phase species onto very cold (10-30 K) grains leads to the formation of saturated molecules and, because of the predominance of hydrogen, saturated hydrides of abundant elements (O, C, N) will be of prime importance for dirty ices formation as proposed as early as 1946 by Oort and van de Hulst (1946). Gas phase reactions will make up for important amounts of CO and associated molecules ($H_2CO$ and $CH_3OH$ in particular). These molecules will also end-up in icy mantles. As a matter of fact, various irradiation agents (photons and energetic particles) will produce an extremely complex solid-state chemistry on the grain surfaces and within the bulk of the ices, a kind of radical-radical chemistry (Butscher et al, 2016) that will greatly increase the molecular complexity in the solid state because the grains themselves do offer protection to newly formed and increasingly complex species. This leads to a true molecular complexity that remains however difficult to observe in the infrared (Raunier et al, 2004). We can note here that the next generation of space telescopes, including the JWST (2021), will greatly benefit to the studies of interstellar ices and to observations of organic residues on the grains, residues that are so easily made in the laboratory simulations.

As described in Le Sergeant d'Hendecourt (2011), the global evolution of the chemistry of the interstellar medium can be sketched in the following manner, as represented in Figure 1: in the evolution of our Galaxy, nucleosynthesis in stars produces the heavy elements (above H) and supernovae explosions do enrich the interstellar medium in all the elements that belong to the Mendeleev table. This will progressively determine a hierarchy in relative elemental abundances that is generically known as the "cosmic abundances" of the elements (Tolstikhin and Kramers, 2008). Generally speaking, H and He, elements that are only gaseous, dominate over the first group of elements that can form (or not) solids, O, C, N, S and, to a lesser extent, P. These elements account, in total, for around $10^{-3}$ the amount of hydrogen, which, as a direct consequence, explains why the gas to dust ratio in mass in molecular clouds takes the canonical value around 100. The elements Si, Mg, Al, and Fe will account altogether for $10^{-4}$ H. All the other elements can be considered as trace elements as far as the astrochemistry we consider here goes. Stars like our Sun display these cosmic abundances in their photospheres and, although the Galaxy shows some local differences in elemental abundances, essentially from its center to its periphery, for our purpose here, this description is sufficient.

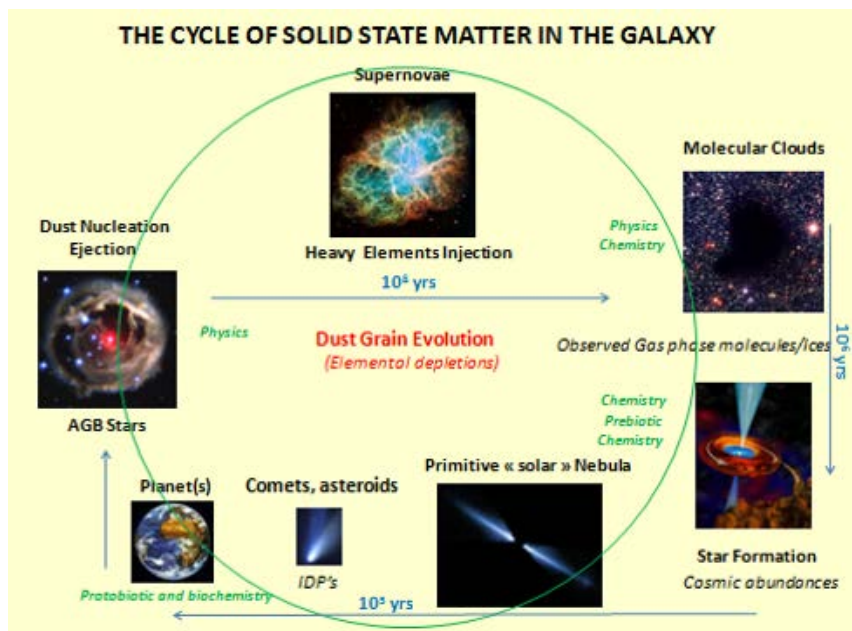

Figure 1: A global view of the chemical evolution of our Galaxy with emphasis on the solid state materials, extraterrestrial grains and ices. Typical evolutionary times are indicated in this continuous process (Le Sergeant d'Hendecourt, 2011)

Stars at the end of their lives, start losing their envelopes which then leave the dying red giant. Expansion cooling off the flow will allow, as in a jet nozzle, nucleation and condensation of mineral grains essentially made of refractory materials like minerals such as silicates (e.g. $MgSiO_3$ and many other ones), see Henning (2003). These heavy elements being refractory, make permanent solids and, in contrary to the volatile species, are lost for the gas phase chemistry. Around the dying star, these lighter elements, the CHNOPS, are cosmically more abundant but do not necessarily condense in refractory solids. These elements may form some gas phase molecules such as $H_2O$ and CO but ultimately these species reach the diffuse ISM (far from their star) and dissociate. Since these elements are not entirely lost in the make-up of the solid grains, they are observed to be much less depleted in the diffuse ISM than the elements like Fe, Mg, Si, Al that are fully depleted by at least by two orders of magnitude (Palme and Jones, 2003), an observation which was interpreted very late in the development of radio astronomy (Greenberg, 1974). Entering the molecular clouds, the chemistry will start with these gas phase elements and this is the reason why organic chemistry fully dominates the interstellar medium chemical evolution. So, cosmic abundances *and* thermodynamic properties of the nascent molecular solids in stars envelopes do favor the rich organic chemistry observed in the gas phase of dense molecularnclouds. Separation of the CHNOPS elements from the refractory mineral elements that give the bulk composition of planets like Earth (a rocky planet) is a starting condition for the prevalence of organic chemistry in astrochemistry. One step further, as we shall describe, the CHNOPS will greatly contribute, through the making and evolution of the ices, to the formation of organic matter upon energetic processes, a matter that will display many important biologically relevant molecules. The fact that amino acids, sugars and the likes do contain only the CHNOPS directly may indeed reflect the importance of the global astrochemistry of the icy materials in extraterrestrial settings, and, as a direct consequence, the importance of this material for further prebiotic processes at the surfaces of telluric planets.

## II  TOWARD THE ORGANIC RESIDUE: LABORATORY EXPERIMENTS

Since the 1980's in the group of Prof. J. Mayo Greenberg in Leiden, experiments were designed to test the importance of cosmic ices in molecular clouds. Basically, these experiments are using classical matrix isolation techniques and their astrophysical potential is explained in d'Hendecourt and Dartois (2001). Such experiments are quite generic and performed by many different laboratories throughout the world. Our experimental set-up is named MICMOC (Matière Interstellaire et Cométaire; Molécules Organiques Complexes). It has been conceived at IAS (Orsay) in 2003 and since 2017 is in use at PIIM (Marseille). Some recent description of our own set-up can be found in de Marcellus et al (2017) and in Fresneau et al (2017). A gas mixture is prepared in a stainless steel evacuated line. The gas is then slowly deposited onto a cryo-cooled window transparent to infrared radiation (usually an $MgF_2$ window), cooled to the desired temperature (10-30 K or, as in our case, 77 K), as illustrated in Figure 2. A hydrogen microwave discharge plasma lamp is used to provide a strong ultraviolet flux of photons mostly in the Lyman $\alpha$ range (but not only). Note that these experiments were originally built-up to explain the infrared astronomical features observed by the SWS spectrometer on board the ISO satellite, a goal which was readily achieved with the first results from this instrument as evidenced in d'Hendecourt et al (1996).

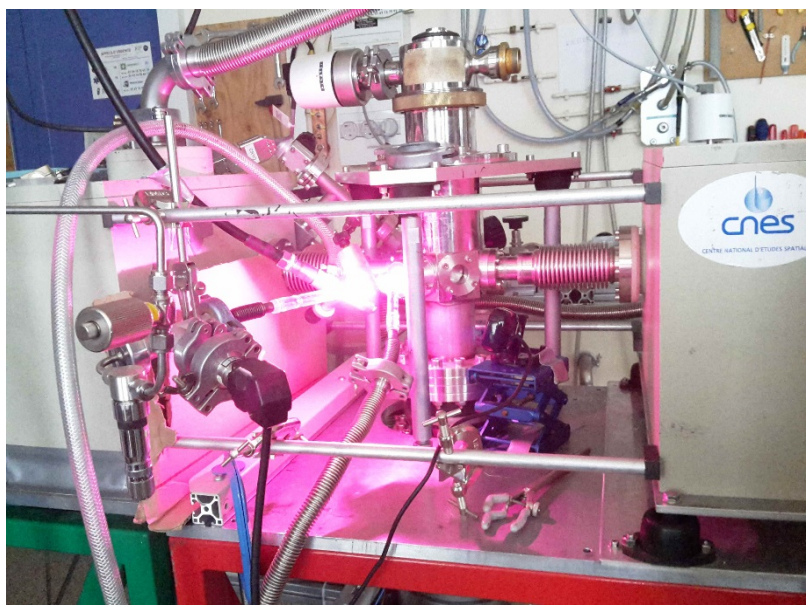

Figure 2: The MICMOC experiment at IAS (now at PIIM, Marseille). An open-cycle cryostat (vertical), cools an $MgF_2$ window at its center. The UV irradiation is obtained from a plasma microwave discharge $H_2$ lamp that produces this pinky color (actually mostly H$\alpha$, reminiscent of the colors of many nebulae in the galactic sky) but the irradiation (inside the cryostat) is mostly in the Lyman $\alpha$ band.

Our philosophy is to provide a plausible template of an astrophysical sample that may be consistent with the composition of primitive carbonaceous chondrites. The approach is essentially *phenomenological* in the sense that, as soon as solid state radiation chemistry starts to play a role, the complexity of the reaction network to be considered goes far beyond the possibility offered by any necessarily limited chemical network model. The photo- and thermo-chemistry of only three solid state molecules (as in our case $H_2O$, $CH_3OH$ and $NH_3$) paves the way to the build-up of thousands of byproducts so that the complexity of the obtained organic residue cannot be understood in a mechanistic and fully reductionist way. The gas phase mixture, almost always the same for simplicity in the analytical procedures, is deposited onto the cold window at 77 K and simultaneously photolyzed by the UV lamp in order to obtain a full in depth penetration of the UV radiation within the totality of the sample thickness. Our equipment (deposition rate and UV lamp flux) is calibrated so that the radiation dose imparted to the whole sample does not exceed 1 to 3 Lyman α photons per deposited molecule. A temperature of 77 K was chosen in order to increase the diffusion of the reactants during the irradiation step and thus increase the efficiency of organic matter formation. Anyhow, one must also remember that the sample is recovered, only after the evaporation of most of the ice and at room temperature (300 K) for analytical purposes. Although not discussed further here, we note that variations in some parameters (ice composition, ultraviolet dose or energy source and temperature) have been described by ourselves (Fresneau et al, 2017) and others (see for example Munoz-Caro et al, 2014). The final residue proves not to be very sensitive in composition to these parameters as far as the initial elemental relative contents remain similar, a fact that shows indeed that the final result displays a chemistry with not much selectivity and essentially assisted by radical-radicals/molecules recombination (d'Hendecourt et al. 1982). Finally, an increase in the total irradiation dose up to the 10 keV range per deposited molecule, irradiating further the residue under vacuum and at room temperature, leads to an insoluble organic component that resembles the insoluble organic matter known as IOM and present in primitive chondrites. This shows that our simulations are able to provide a realistic template of an extraterrestrial sample, analogous in many aspects to the organic matter present in primitive chondritic meteorites (de Marcellus et al, 2017).

A picture of a typical residue is shown in Figure 3. Droplets of organic material are observed under the microscope and present an infrared spectrum that is characteristic of classical functional groups that can be found in organic materials such as the soluble organic ones that are currently observed (Sephton, 2002, Ruf et al, 2018) in carbonaceous meteorites with the presence of ketones, ethers, esters, carboxylic acids, aliphatic hydrocarbons, amines, amides…. It is important to note that the whole sample, in the case of the low irradiation regime, is *fully soluble* in solvents such as water or methanol and can be easily extracted for further analysis using efficient analytical techniques such as the ones offered by Gas phase Chromatography and Mass Spectrometry techniques (GC-MS). Altogether, between 50 and 300 μ-grams of material are recovered after a few weeks of experiments, as the growth of this material is a rather slow process. We also assume that this highly complex and soluble organic residue may be used for prebiotic evolution, as solubility is a good and probably essential criterion for this purpose.

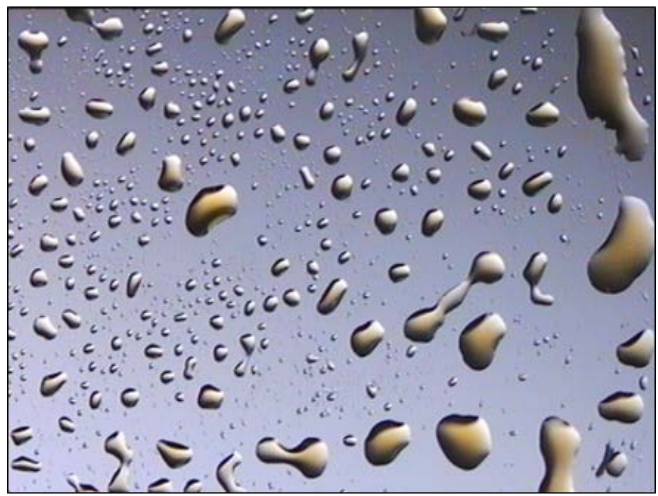

Figure 3: A typical residue on the MgF$_2$ window under the microscope (x60, white light). Droplets of organic molecular oily residue are observed. They represent a total mass of soluble matter around 100 µg. The yellow color is always observed on these materials, nicknamed "yellow stuff" around 1980 by J. Mayo Greenberg. Picture size is around 1x1 mm$^2$ (Modica, 2014).

## 2.1 Non targeted vs targeted searches for specific molecules

Two general classes of analytical procedures can be used to study these organic residues. The first method is a *non-targeted* search in which one hopes to obtain an immediate photography of "all" the compounds present in the residues without any precise assumption of its chemical composition regarding its potential prebioticity as well as without any prejudice on this composition from the experimenter's side, an original view that is discussed in Ruf et al (2018). By "all" we wish to emphasize that the recovery and assignment of the molecular compounds may suffer from some bias in their detection. The most used method here will be very high resolution mass spectrometry ($\delta$m/m over 10$^6$) using FT-ICR, a method that has proven quite efficient for some revolutionary work on the Murchison meteorite (Schmitt-Kopplin et al, 2010), with the detection of more than 15000 compounds and a countless number of isomeric species. Similar methods using less resolving instruments (an Orbitrap) on the *same organic residues* from ices have also revealed a very large number of molecular species including the presence of macromolecular structures up to 2000 daltons (Danger et al., 2013). Quite convincing similarities with the organic matter from meteorites were also discussed by Danger et al. (2016). These results are nevertheless out of the scope of this paper and will not be discussed further. The second analytical method involves a *targeted* search for specific molecules, and, for the purpose of this work, amino acids and sugars were investigated in the residues.

## 2.2 Amino acids in laboratory organic residues

Prior to 2002, the build-up of residues after warm-up at room temperature of the deposition window was already considered and some analytical tools involving GC-MS methods were tried. Essentially, in the group of J.M. Greenberg in Leiden (The Netherlands) and collaborators, the analysis led to the detection of many chemical compounds out of which carboxylic acids could be pointed out (Briggs et al, 1992). However, the detection of amino acids in large numbers (up to 16) was achieved for the first time in 2002 (Munoz-Caro et al, 2002), in a joint effort from both groups from J.M. Greenberg and U.J. Meierhenrich, at this time in Orléans (France). In the same issue of Nature, an independent group of NASA Ames scientists arrived at a similar conclusion (Bernstein et al, 2002) albeit with less efficiency (only 4 amino acids were detected there). Actually, the chemical treatment used to enhance the detection capabilities in an a priori unknown sample, involves methods that were first developed to extract amino acids from meteoritic materials, in particular from the famed Murchison one. The analytical protocol is fully described in a more recent work, on the same residues (Meinert et al, 2012): the sample is first extracted from the window using methanol and taking advantage of the noted complete solubility of the residue in organic solvents. The residue is then hydrolyzed in HCl in order to free the amino acids from their possible arrangement in macromolecular species. Such a technique allows for an increase in the free amino acid content. Before passing through the GC-MS, actually a bi-dimensional GCxGC-MS in the experiments of Meinert et al. (2012), the amino acids follow a classical procedure of derivatization which allows an easier detection of the heavier and less volatiles compounds by giving them a higher volatility. The list of detected amino acids (20 different ones) with their relative abundances can be found in Meinert et al (2012).

**2.3 Aldehydes and sugars**

If amino acids have always received the prime attention as far as extraterrestrial biologically relevant compounds are concerned, and for obvious reasons starting with the historical experiment of S. Miller (1953). It remains however quite self-evident that other molecules may be of major interest to potentially constitute an efficient reservoir of such molecules in extraterrestrial materials (meteorites) and considering a scenario involving the exogeneous delivery of organic matter onto the primitive Earth. In our case, using *exactly the same kind of samples*, we decided to look for aldehydes and sugars. They are indeed made using the same MICMOC experiment, exactly the same initial molecular gas phase mixture, the same time of deposition and irradiation dose and the same extraction procedure for the recovery of the residues. Further analytical procedures focus on the search for sugars. Obviously, the detailed analytical protocol is then different, does not involve acid hydrolysis that would destroy the aldehydes and makes use of a different derivatization process as explained in de Marcellus, Meinert et al (2015) and Meinert et al (2016). A certain number (8) of aldehydes including glycolaldehyde and glyceraldehyde were first detected, quickly followed by the finding of a suite of sugars and related molecules such as sugar alcohols and acids.

These sugars do contain many different aldopentoses and in particular, ribose, the backbone sugar of the RNA molecule. A list of the detected aldehydes is given in de Marcellus, Meinert et al. (2015), and sugars and related molecules can be found in Meinert et al. (2016). Note that, although glycolaldehyde is often considered as a sugar by radio-astronomers, glyceraldehyde is in fact the first ($C_3$) sugar of the -ose family and is indeed a *chiral* species on which some of our efforts remain concentrated. Remarkably, glycolaldehyde is now detected in radio astronomy, thus as a gas phase species, in many locations such as low mass protostars (Jorgensen et al., 2012) and even in the coma of a comet (Biver et al (2015). Note here that the non-detection of glyceraldehyde in these same astronomical objects may well reflect the *extremely biased* detection of any astronomical molecule radio detection due to many factors that include (i) a more complex radio spectrum and thus emission lines distributed over a much larger partition function, (ii) a lower abundance of the more complex molecule (a factor of roughly ten in our own experiments) as it is *always* the case in radio astronomy, and (iii) a much lower vapor pressure for vaporization from the ice to the gas phase that significantly precludes a detection in this phase. Any observations in gas phase astrochemistry is strongly biased and this simple fact is quite often overlooked.

Finally, before we close the case for the presence of sugars in these residues, one must emphasize that, to our surprise, the total amount of all the sugars (and related molecules) detected in these experiments is very large as it amounts to 3 to 4% in mass of the solid sample. Together with the high number of various sugars detected (trioses, tetroses and pentoses out of which ribose is the most remarkable) a tentative explanation was proposed in Meinert et al (2016, 2017), for this very attractive characteristics of our samples. We indeed postulated that the sugars were actually formed during the sublimation of the irradiated ices by a kind of formose reaction, a quite well known and classical reaction (Butlerow, 1861; Breslow, 1959) which involves the polymerization of formaldehyde trapped in the ice and detected in-situ via infrared spectroscopy, leading to the formation of glycolaldehyde which, following an autocatalytic reaction, leads to this almost explosive burst of sugars. In the laboratory, the formose reaction is encouraged by a strong base agent involving calcium hydroxide that is not present in our experiment. However, in such matrices irradiated by a strong ultraviolet flux, the presence of solvated electrons in the ice (Eiben and Taub, 1965) may play the role of a catalyst for this "formose-like" reaction. Such a reaction, if present in evaporating samples, as well as in cometary materials, may well have some strong consequences in astrobiology because it might be a very generic process in extraterrestrial ice chemistry and produces *at the same location and at the same time*, a wealth of amino acids *and* sugars.

## III SYMMETRY BREAKING IN AMINO ACIDS IN ASTROPHYSICAL ICES

As well known in the astro- and cosmo-chemical communities, primitive carbonaceous chondrites, which display a large suite of organic molecules including those detected in the experiments described above, do clearly show enantiomeric excesses (e.e.'s), mostly of the L form for certain chiral amino acids. It is not the purpose of this paper to discuss in detail this peculiar aspect offered by these primitive materials. The reader may refer to recent reviews on the subject by Myrgorodska et al., (2014) and Burton and Berger, (2018). As known, in living systems, proteins are made uniquely of L amino acids, while RNA and DNA are made only of D sugars. This property is known as homochirality. Conversely, in non-biotic processes, as in our experiments, a suite of chiral amino acids and sugars are produced as strict racemic mixtures. The origin of homochirality is unknown as well as the origin of the excesses of L amino acids detected in some meteorites. Since we postulated that biologically relevant molecules can be synthetized following a plausible astrophysical scenario, it seemed logical to our team to perform experiments similar to the ones presented above, but this time using *ultraviolet circularly polarized light* (hereafter UV-CPL). In this approach, from an astrophysical point of view, the true incentive of this work, comes from the fact that circularly polarized light is indeed observed, on large spatial scales in many protostellar environments (see for example Kwon et al. 2014), exactly in the *same locations* where extraterrestrial ice photoprocessing takes place.

### 3.1 A realistic astrophysical chiral agent: circularly polarized light

To remain coherent with the astrophysical motivation of our experiments, we concentrate here only on the possible role of UV-CPL on the build-up of enantiomeric excesses in chiral complex molecules that are *in-situ synthetized* in our icy samples. Although many and very diverse possibilities have been described to conceivably generate symmetry breaking in various chiral media (Meierhenrich, 2010), we limit ourselves here to a scenario involving processes that can be *supported by some observations.* Indeed, among the mechanisms thought to generate significant e.e.'s in astrophysical environments, UV-CPL is one of the most frequently described (see Bailey 2001 and references therein). As a matter of fact, it is a phenomenon suggested a long time ago by J. H. Van't Hoff as far back as 1897. However, it has never been experimentally proven in a self-consistent manner, one which examines s*imultaneously* the build-up of a reasonably complex organic chiral molecular pool, as in our case amino acids, together with an effective symmetry breaking process. UV-CPL being susceptible to induce photochemistry in ices, just as unpolarized UV photons do, we tested this scenario in an identical ice simulation (the Chiral-MICMOC experiment) but using UV-CPL to drive the photochemistry of cosmic ice analogs and search for the generation of chiral species with s*ignificant* e.e.'s, an approach proposed by Greenberg et al (1994). Note an interesting peculiarity here is that the initial ice mixtures used, the same as described above, are made of $H_2O$, $CH_3OH$ and $NH_3$ and that these molecules are clearly *achiral*. In other words, these initial simple molecules are insensitive to the helicity orientation (L or R) of the UV irradiation light. Previously attempted experiments, some performed by our team, provided inconclusive results, either due to intrinsic limitations in the photon flux and/or to the detection limits of the analytical methods used (Nuevo et al. 2006; Takano et al. 2007), or because they were not readily applicable to a realistic astrophysical situation (Takano et al, 2007), a condition that is indeed our top priority. However, as clearly shown in Nuevo et al (2006), they opened the route to the final success of this approach.

**3.2 The SOLEIL experiments**

We have performed four different experimental campaigns at the SOLEIL synchrotron from 2009 to 2012. The experimental setup, which has been presented above, was adapted to be mounted on the vacuum UV DESIRS beamline at the SOLEIL synchrotron facility (France). The SOLEIL experiments and first results are described in de Marcellus et al (2011) and more recent and complete results are presented in Modica et al (2014) together with a proposed astrophysical context. The undulator feeding the DESIRS beamline provides a high quality VUV-CPL beam (97% pol.) of any chosen helicity, i.e., right (R) or left (L), as well as linearly polarized light (LPL). This radiation, has a typical 7% spectral bandwidth and an integrated flux of about $1 \times 10^{15}$ photons cm$^{-2}$ s$^{-1}$ at the sample position. Interestingly, this characteristic is quite similar in photon flux to the one, in the Lyman α range, given by our own lamp. For the first set of experiments, we selected an irradiation photon energy of 6.64 eV, which is close to the maximum intensity of the circular dichroism (CD) spectrum ($n$–$\pi*$ transition) of most $\alpha$-hydrogenated amino acids measured in the amorphous solid state (Meierhenrich 2010). *Empirically,* our goal was to search for enantioselective synthesis or destruction of chiral amino acids triggered by UV-CPL. Ice mixtures of composition of $H_2O$:$CH_3OH$:$NH_3$ (2:1:1), were deposited onto a cold ($T$ = 77 K) IR-transparent window in order to monitor the evolution of the composition and thickness of the ice, and to correlate these parameters with the number of synchrotron delivered photons irradiating the ice samples during the experiments. As in the original experiments, the use of a higher temperature (77 K), rather than expected temperatures for interstellar ices (10–30 K), was decided to enhance diffusivity and recombination of photoproducts (radicals) within the ices. As previously discussed, since the complete cycle of inter and circumstellar grain evolution comprises cycles through warmer regions (hot cores) in which grain temperatures may rise significantly (200 K or larger), astronomical organic residues should be produced via pathways similar to those simulated in our experiments. Finally we may note that, on the Earth, meteorites picked-up for analytical work are at a temperature similar to our room temperature in the laboratory (300 K) thus validating our simplified and empirical approach, keeping in mind that in analyzed meteorites such as Murchison, internal temperatures remain well below 300 K during the fall and then get heated on Earth to the local temperature.

*3.2 1 The first set of experiments: enantiomeric excesses in alanine (2009)*

The ability to cleanly separate both left and right enantiomers and also separate them from overlapping peaks by applying two-dimensional GC techniques, allows for the detection of minute but significant amounts of e.e.'s. In this first test experiment, only the simplest chiral amino acid, alanine, had an abundance that allowed for a precise measurement of these minute e.e.'s. The measured e.e.'s of alanine for each of the three different UV light polarization regimes explored is given in Figure 4 (pink panel, year 2009), and are e.e.L = −1.34% (±0.40%) and + 0.71% (± 0.30%) for the R- and L-CPL samples, respectively. The LPL sample provides, as expected, an e.e. very close to zero (−0.04% ±0.42%), confirming that the excesses are indeed due to the UV light polarization state. Note that the error bars are 3$\sigma$, giving a confidence level larger than 99.7% for these measurements. Although the photon-induced e.e. values switch sign with the photon helicity as expected, the absolute e.e. for the R-CPL sample is ∼2 times larger than for L-CPL. Actually, the R-CPL sample was built up with roughly twice as many photons per deposited molecule ($N$phot.molec−1) as the L-CPL sample. To first order, this ratio turns out to be comparable to the ratio of the measured absolute e.e.'s, so that the absolute enantiomeric excesses appear to be proportional, in our experiments, to the number of "chiral" photons irradiating the samples. The enantioenriched formation of stereogenic centers in the organic matter is clearly favored by a higher ratio of CP photons over deposited *achiral* ices. The net result of this experiment is that the UV-CPL irradiation of a plausible cosmic ice mixture consisting initially of only achiral molecules induces the production of *significant* e.e.'s of photon helicity-dependent signs for alanine, depending on the helicity of the CPL. Contamination issues are resolved by the use of $^{13}$C-labeled methanol, the only source of carbon here, and the measurements imply $^{13}$C alanine only to avoid for any contamination during manipulation of the samples. Note that the absolute value of the maximum excess (1.34%) is compatible with the e.e.L measured for alanine in the Murchison meteorite (1.2%), as well as for some other meteoritic $\alpha$-hydrogenated amino acids (Cronin & Pizzarello 1999). For the campaign of 2010, in order to investigate whether the chiral properties arise during the irradiation of the ice or of the residue, three samples were prepared. R CPL irradiation was used at two different stages: ice and the residue. A third sample was irradiated by LPL only during the ice stage. The results show that the induced e.e is independent on the sample stage during CPL irradiation (Figure 4, pink panel, 2010). Finally, note that, on Figure 4, different sets of data do appear, corresponding to SOLEIL campaigns from 2009 (first results, de Marcellus et al, 2011), to 2012. These latter results will be discussed in the next section.

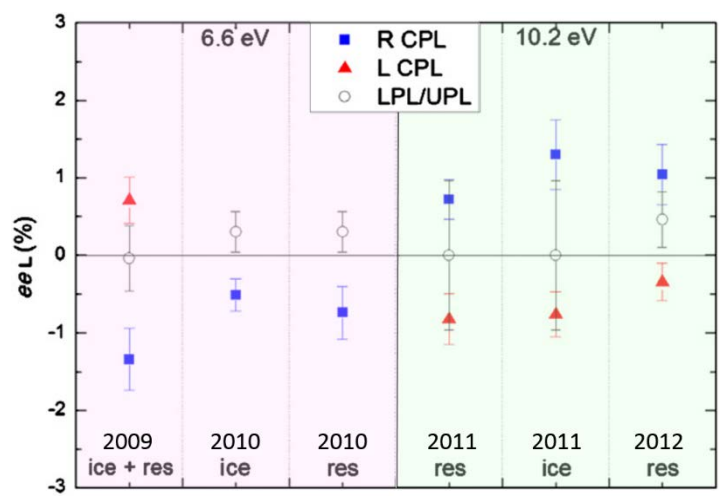

Figure 4: Measured *eeL* in $^{13}$C alanine for the different synchrotron sessions over four years. The pink panel includes experiments conducted at 6.6 eV, the green panel those at 10.2 eV. Blue squares represent *eeL* induced by R CPL, red triangles those by L CPL, and white circles those after LPL or UPL. Note that the sign of the *eeL* depends on the helicity and the energy of CPL and do not depend on the stage of the sample (residue or ice). Figure adapted from Modica et al (2014).

*3.2.2 The second set of experiments: e.e's of the same sign in five amino acids (2011 – 2012)*

In order to validate and extend these first results, we decided, in a second series of experiments on the same beamline, to vary somewhat the experimental protocol. The irradiation wavelength was chosen this time to be Lyman α only (10.2 eV), essentially because irradiating at different wavelengths is time consuming and practically not feasible owing to strict beam time allocation limits. Since the optical dichroic properties of amino acids, and in particular their CD spectra (Meierhenrich, 2010), are not known in this wavelength region, our experiment is exploratory. It is indeed a priori impossible to hint at the results prior to perform the experiments when, on alanine, the wavelength was adequately chosen to maximize the effect of the CPL. In 2011, the experiments were performed at conditions that allowed investigating the effects of the irradiation by the two CPL helicities (R and L) together with the stage (ice or residue) at which the samples were irradiated by CPL; one sample was irradiated by LPL at the ice stage. The amount of residues produced was small (<100 μg) and comparable to that obtained during the previous series of experiments. Since the experiment is a passive one regarding the irradiation dose which requires for a long sample build-up time on the beamline, in 2012 we decided to first produce our samples in the original facility at IAS with the Lyman α unpolarized light, provided by our own vacuum UV $H_2$ microwave discharge lamp (see Figure 2). In this last series of experiments, three samples of increased quantity of organic residues (>100 μg), compared to the ones of the former experimental campaigns, were produced. The samples produced were then transported in a specific vacuum chamber, to avoid oxidation from the air, and mounted at the extremity of the DESIRS beamline to be irradiated by the UV-CPL, actually this time only at room temperature and in high vacuum. This procedure minimizes the beam time to be requested from the SOLEIL time allocation committee and allows for the make-up of somewhat larger samples (usually a factor of three) that are easier to produce at IAS and thus the possibility to detect and evaluate e.e's on more amino acids. This procedure together with the results are described in Modica et al (2014).

The main results are displayed on Figure 5. The green background represents the CPL energy fixed at 10.2 eV (Lyα), as on Figure 4. Five pairs of L and D chiral amino acids could be detected and their ee's quantified. Interestingly, at Lyα, they do show small e.e's of the same sign when the same CPL helicity is used for the irradiation. This behavior confirms the one reported from the first results on alanine, but this time it is extended to a set of five distinct pairs of amino acids. As we will explain thereafter, it is indeed important to obtain, under a certain illuminating light, the *same helicity* (L or D) for the detected amino acids otherwise, one of the main peculiarities from e.e's in meteorites would not be fit by contradicting results on the e.e's signs.

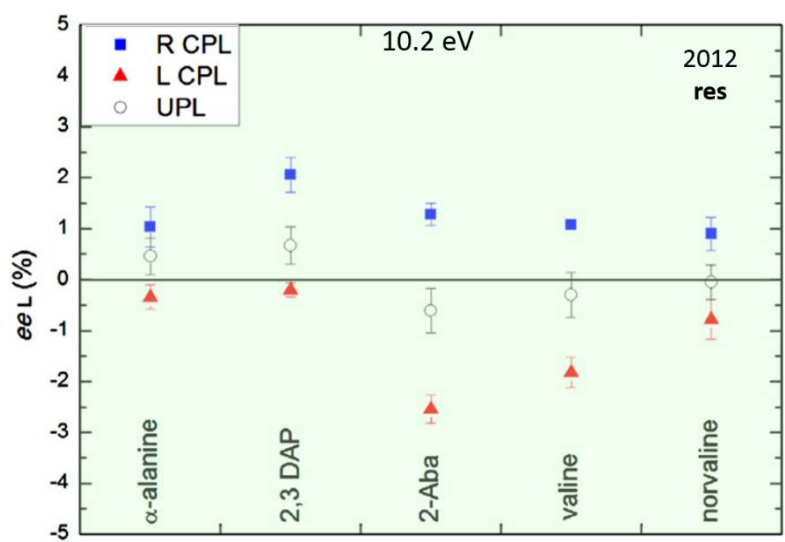

Figure 5: Enantiomeric excesses obtained for five amino acids in the last SOLEIL campaign (2012). Three enlarged residues were irradiated by CPL at 10.2 eV (Lyman α). Blue squares represent *eeL* induced by R CPL, red triangles those by L CPL, and white circles those after LPL or UPL. Note that the *eeL* are of the same sign in all five amino acids for a given helicity of CPL. Figure adapted from Modica et al, (2014).

*3.2.3 Discussion on the results*

Higher enantiomeric excesses, such as those measured for isovaline in Murchison (up to 18.5%, see Pizzarello et al. 2003 and Glavin & Dworkin 2009), cannot be explained solely by the asymmetric effect of UV-CPL. Amplification of a small initial enantiomeric excess, such as those obtained in our experiment by aqueous alteration on the parent body, has been proposed (Glavin & Dworkin 2009) but for which no precise explanation does exist yet. These excesses can later be amplified by catalytic processes to reach a nearly homochiral state (Bonner 1991), a phenomenon that will not be discussed here.

Starting from an achiral ice mixtures, photo- and thermochemistry leads to the formation of complex residues made of essentially stereogenic centers and chiral molecules. Once started, these materials do see the "chiral" photons from the UV-CPL used in the experiment and some selection of one enantiomer is thus favored. Our experiments show that this selection operates as soon as some UV-CPL coexist with non-polarized light. The effects do add together, both making chiral molecules when the polarized light select enantiomers. Actually, although somewhat phenomenological at this stage, this effect is indeed expected in the interstellar medium as discussed in the next section.

## IV A PLAUSIBLE ASTROPHYSICAL SCENARIO

The astrophysical framework, the making-up of organic materials from cosmic ice photo- and thermo-chemistry, has already been presented in §I of this paper. In the particular context of the build-up of enantiomeric excesses linked to these photo-chemical processes, our present simulations may be applied to the outer solar nebula evolution, where icy grains may have been irradiated by an external source of UV-CPL. The proposed scenario is described in detail in Modica et al (2014) and we will just present here the main points of this scenario.

We first adopt the model of Ciesla and Sanford (2012): during the early phases of the protoplanetary disk evolution, icy grains driven by disk instabilities moved along radial and vertical motions to be accreted onto the young Sun. Part of them however survived and, in this model, the path of these remaining grains was calculated through the disk, showing that they could have been exposed to substantial *external* UV irradiation (~50 photons per ice molecule) during a lifetime of $10^6$ yrs. Second, UV-CPL could have irradiated such an environment: the best candidates for CPL sources are reflection nebulae in star-forming regions associated with a dominant high-mass young stellar object (YSO). High degrees of CPL have indeed been observed in the near-IR in reflection nebulae such as OMC-1, at levels of 17% (Bailey et al. 1998; Fukue et al. 2010), and more recently in NGC 6334, at levels of 22% (Kwon et al. 2013). A discussion on the polarization mechanisms based on light scattering by non-spherical grains aligned by the star's magnetic field, or on dichroic extinction of light that has been previously linearly polarized by scattering on aligned non-spherical grains can be found in Chrysostomou et al. (2000) and in Kwon et al. (2013). Active sites of star-forming regions present a too high dust obscuration for direct observation of UV-CPL. However, detailed models for aligned grains with small axis ratios have shown that CPL can also be present in the UV range with similar high levels of polarization as those observed in the IR (Bailey et al. 1998).These CPL regions present a quadrupolar-shape pattern of polarization where each quadrant of a single-handedness covers a wide area (Fukue et al. 2009). Indeed, the extent of CPL in OMC-1 is about 0.4 pc (Fukue et al. 2010), while in NGC 6334 it is about 0.65 pc (Kwon et al. 2013), which is hundreds of times the size of most forming planetary systems, including ours. If our solar system formed in a similar star-forming region, it may have been irradiated by CPL of a *single handedness*. Most low-mass stars like our Sun are indeed believed to be born alongside high-mass stars, as it is observed in the case of OMC-1, where high- and low-mass stars are forming together. This hypothesis is supported by the presence of short half-life radionuclides such as $^{60}$Fe and $^{26}$Al in primitive meteorites (Gounelle & Meynet, 2012). Under these conditions, this irradiation may well have been circularly polarized, with a single helicity. The main amount of CPL would have been received outward and at high latitude in the disk, where lower densities prevented the flux from being too strongly attenuated as compared to the mid-plane. The episodic exposures to this increased photon flux would have triggered the double effect of asymmetrically photoprocessing and heating of the ices. These processes may have converted at least 5% of the ices into organic compounds (Ciesla and Sandford, 2012), including chiral molecules such as amino acids and/or their precursors. Along with the synthesis of interstellar/circumstellar organic compounds, the asymmetric photoreactions triggered by CPL may have *simultaneously* induced an *ee* in any chiral molecule. As suggested by our current results, this action may have occurred at both stages in which grains were covered by ices and/or in which they were already covered by a refractory residue of organic compounds, thus maximizing the possibility to accumulate a significant *ee* before the accretion into larger grains, planetesimals, and finally comets and asteroids.

Regarding the pertinence to asymmetric photoprocesses, we point out our choice to consider the VUV spectral range at higher energies (10.2 eV) than those typically considered so far. Indeed, according to the model of Robitaille et al. (2006), YSOs of higher masses (∼20$M$sun) do present a spectral energy distribution (SED) favoring high-energy photons at the earliest stages of their evolution. Over time, this emission becomes dominant and its maximum is moving to higher energies. At stage II of the protostellar evolution, which is associated with the presence of an optically thick disk and an infalling envelope, the emission is peaked around 10.4 eV (~0.12 $\mu$m), the one chosen in our synchrotron experiments, there again giving some credit to this tentative scenario. Finally, a last word in favor of this hypothesis: recent observations of numerous star forming regions by Kwon et al. (2014) do show that IR-CPL (and by extension local UV-CPL) are a general characteristic feature of star forming regions and certainly not an exception. Thus if CPL is able to generate asymmetry in chiral astronomical molecules as demonstrated here, this fact may be a very general trend in many planetary systems in which chiral excesses would then be present for further prebiotic chemistry.

**V CONCLUSIONS**

Dedicated laboratory experiments that simulate the photo- and thermo-chemical evolution of cosmic ices do easily produce a wealth of molecules often called "molecular bricks of life" as in our case, amino acids and sugars. The MICMOC experiment, a non-directed methodology, was designed to follow the complex evolution of molecular ices and their transformation into organic materials. It has been then been followed by the chiral-MICMOC approach, in which, with the use of UV-CPL from the synchrotron SOLEIL, we managed to produce measurable and significant *enantiomeric excesses of the same sign for five amino acids out* of an original achiral molecular mixture of simple molecules in dirty ices. Moreover, the produced molecular organic complexity has been shown to present numerous similarities with the soluble organic matter in carbonaceous meteorites.

Obtaining enantiomeric excesses has proven to be feasible in the laboratory using a circularly polarized light, UV-CPL, which is indeed present and observed, albeit only in the infrared, in many star forming regions where protoplanetary disks form and evolve. These observations, together with sophisticated models of the solar nebula and its outside irradiation with UV-CPL show with some strength, that indeed initial chiral excesses in some molecular bricks of life such as amino acids may be produced at the onset of the protoplanetary disk phase that became later the solar system. One must emphasize though, that due to the quadrupolar structure mentioned in the young stellar objects where CPL is observed, that other regions of star formation may well display a different helicity than the one given to the solar system and thus an opposite sign for the amino acids of any living system if present there. The delivery of organic matter (and water) by impacts of cosmic materials is one of the most favored hypothesis to feed up the prebiotic chemistry at the surface of the early Earth. The genericity of this process, if not fully established yet, is a plausible hypothesis following recent observations of new born planetary systems where organics and water are indeed largely observed.

Finally, it must be stressed that the L form of the amino acids detected in primitive meteorites for some amino acids do match the L homochirality form for amino acids in the living world. Although not a definite proof, this suggests a true molecular connection between these meteoritic amino acids and prebiotic chemistry. One may ask then the same question for the sugars that are indeed also produced in these ices.

**Acknowledgment**

Louis d'Hendecourt is grateful toward the French space agency (CNES) and more particularly Dr. Michel Viso for the constant moral and financial support of the MICMOC and Chiral-MICMOC experiments described in this article.